\documentclass{article}

\usepackage{arxiv}

\usepackage[utf8]{inputenc} 
\usepackage[T1]{fontenc}    
\usepackage{hyperref}       
\usepackage{url}            
\usepackage{booktabs}       
\usepackage{amsfonts}       
\usepackage{nicefrac}       
\usepackage{microtype}      
\usepackage{lipsum}		
\usepackage{lineno}
\usepackage{graphicx}
\title{On the Co-movement of Crude, Gold Prices and Stock Index in Indian Market}

\author{
  Abhibasu Sen \\
  Department of Mathematics, \\
  Assam University,\\
  Silchar, Assam, India \\
  \texttt{abhibasusen@gmail.com} \\
   \And
 Karabi Dutta Chaudhury \\
  Department of Mathematics,\\
  Assam University,\\
 Silchar, Assam, India\\ 
}

\begin{document}
\maketitle

\begin{abstract}
This non-linear relationship in the joint time-frequency domain has been studied for the Indian National Stock Exchange (NSE) with the international Gold price and WTI Crude Price being converted from Dollar to Indian National Rupee based on that week's closing exchange rate. Though a good correlation was obtained during some period, but as a whole no such cointegration relation can be found out. Using the \textit{Discrete Wavelet Analysis}, the data was decomposed and the presence of Granger Causal relations was tested. Unfortunately no significant relationships are being found. We then studied the \textit{Wavelet Coherence} of the two pairs viz. NSE-Nifty \& Gold and NSE-Nifty \& Crude. For different frequencies, the coherence between the pairs have been studied. At lower frequencies, some relatively good coherence have been found. \\ In this paper, we report for the first time the co-movements between Crude Oil, Gold and Indian Stock Market Index using Wavelet Analysis (both Discrete and Continuous), a technique which is most sophisticated and recent in market analysis. Thus for long term traders they can include gold and/or crude in their portfolio along with NSE-Nifty index in order to decrease the risk(volatility) of the portfolio for Indian Market. But for short term traders, it will not be effective, not to include all the three in their portfolio.
\end{abstract}



\section{Introduction}

In 2013 Wang [1] in his paper on gold as a hedging asset mentioned the statement by de Gaule saying 'Gold has no nationality and is not controlled by governments.'. In the introduction of the same paper, we find a review on how the research
on Gold has been categorised into three groups. As pointed out by the author, the first group discusses the relationship between gold prices and macroeconomic
variables such as exchange rate, interest rate, and income [2,5,42].
The second group discusses the factors affecting gold price 
fluctuations[3,43,44]. The third group discusses the long-term and
short-term relationships between gold prices and the general price index as well
as the effectiveness of gold in avoiding the risks of inflation [4,45,46,47]. Rest of the studies are mostly based on the linear relationship between gold price and macro-economic variables, and use time-series analysis[5,6,42]. Wang [1] also mentioned that the disadvantage of using linear models is that they cannot predict the relationships between variables under different situations. The author also claimed that, there is a non-linear relation between the exchange rates and gold prices based on sophisticated analysis models. Gold
has been a hedge against Dollar and why this hedge has varied, has been discussed in the past[6]. Gold provided a safe haven for investors in the developed markets and has a weak safe haven in the emerging markets[7]. Gold is an excellent protective asset and its value rises due to the
uncertainity involved in the introduction of new financial instruments[8]. There is a relationship between the open interests
of equity futures, light sweet crude oil and Gold futures [9]. Gold works as a safe haven for stocks and not for bonds[10]. Moreover, the hedging strategy works only for a period
of not more than 15 trading days[10]. As a result of which traders and investors from
time to time kept Gold as a hedging tool.
\\ Now, there have been several researches performed on the relation between stock markets, interest rates, inflation  and oil prices.
Several such researches have been done on European markets[11].  Works on GCC (Gulf Cooperation Council) countries have also been done on the oil exporting country's perspective[12]. Increase in oil prices will create a positive impact in the oil exporting countries, while it will create a negative impact on the oil importing countries[13]. Recently various works have been performed on the Chinese stock markets[14-17]. Based on Indian stock markets too, several researches have been performed. Some claim there is a relationship between them[18,49]. But all these researches were done assuming a linear relationship perspective. As a result of which, we are getting a contrasting situation. This drawback is shunned and researches have also been done based on non-linear perspectives too[19-20]. However, the data used by both the groups of researchers had the same periodicity in time.
\\We can see that various researchers have worked on the interlinking between Gold Price, Crude and Indian Stock Market index. There is a long term co-integration between them[21-22].
The above mentioned researches have been done based on Augmented Dickey-Fuller test at levels and 1st differences; as well as Johansen's system Co-integration test. However, now sophisticated methods have been developed, which are being used to study the time-series with different frequencies (or periodicities in time). This technique is known as \textit{Wavelet Analysis} in literature. It basically works on time-scale decomposition, time-frequency analysis etc. This helps in decomposing the frequency in order to understand the market for the investors having different holding periods.
\\There have been many works carried out based on \textit{Wavelet Analysis}. We find a relationship between Crude oil, Gold and Shanghai Stock Exchange(Chinese Stock Market) by using Wavelet Analysis[23]. By Wavelet Analysis method, we find that there is a co-movement between FTSE, DAX and CAC in which FTSE leads[24] (where these three market indices correspond to UK, Germany and France). Based on Indian market too, there have been several studies conducted with the help of wavelet analysis[58,59,60,61]. By \textit{Wavelet Analysis} method, we find there is an important relation between Indian stock market prediction and trading intervals [25]. Studies have been also reported on the relationship between Indian stock market and world stock markets [26]. Wavelet Analysis is used to decompose the data for the prediction of National Stock Exchange (NSE), the Indian Stock Market[27]. Wavelet Analysis was initially used to decompose the data in the frequency domain, after which we get further enhancement in the prediction of NSE-Nifty (Nifty 50 is the index for the first fifty companies of the NSE with respect to market capitalisation) by using artificial intelligence algorithms like SVR and ANN [28]. Post 1997 we also find a high degree of coherence between all the Asian Gold Markets across a group of frequencies [29]. We also find the relationship between the various Asian Stock Exchanges and Bombay Stock Exchange (BSE) using Wavelet Analysis [30]. The scope of risk and portfolio diversification based on Indian Stock Market perspective have also been studied [31]. \\
So we can see that a large number of researches have been performed based on the \textit{Wavelet Analysis} on the two Indian Stock Markets(BSE and NSE). But as such no research has been performed on the study of co-movement between Indian Stock Markets, Gold and Crude Oil Price based on the \textit{Wavelet Analysis}. This paper tries to fill up that gap.
As of 2017, India is the 4th largest importer of crude oil, importing about 60.2 billion dollar oil that is 6.9 percent of the global oil importing market[32]. As a matter of fact, India has the highest private gold holdings in the world, which is about 24,000 metric tons thus surpassing the combined official gold reserves of the United States of America, Germany, Italy, France, China and Russia[33].
\section{Methodology: Description of Analysing Tools}
\subsection{Data Description} The data have been collected from various resources. The NSE-Nifty data have been collected from investing.com[34]. The Gold and WTI crude oil data have been collected from quandl[35,36]. We have also collected the Rupees vs Dollar exchange rate(average of the bid-ask rate of USD/INR) for the corresponding time and multiplied it with the corresponding values of Gold and Crude Oil so as to get the valid comparison between them[37]. We have collected the weekly data from \(5\) November, 1995 to \(8\) August,2018. \subsection{Methodology} We perform the correlation test between Gold, Crude Oil (after conversion to Indian Rupees) and NSE-Nifty Index (Nifty 50 is the index for the first fifty companies of the NSE with respect to market capitalisation) over three different domains. We then perform the Regression analysis. These are for time domain. For frequency domain analysis we first performed the Fourier Transform. For joint time-frequency domain we decomposed the original time series into small wavelets based on Haar Wavelet transformation and perfomed various statistical tests to be discussed subsequently in this literature.
\subsubsection*{Correlation and Regression:} We have the Co-Variance of two variable as the statistical measure, which finds the degree to which the two variables move together; thus capturing the linear relationship between the two variables. It is being defined as:\smallskip 
\begin{equation}\label{}
cov_{XY}= \frac{\sum_{i=1}^{n}{(X_{i}}-{\bar{X}}){(Y_{i}}-{\bar{Y}})}{n-1}
\end{equation}

where the symbols have their usual meaning. \\
Standard deviation is a measure which indicates the amount of variation or dispersion of a set of values. \\
We have the regression model which describes the relationship between two variables (say X and Y) via the equation:

\[Y_{i}= b_{0} + b_{1}X_{i} + \varepsilon_{1}, i=1,\cdots ,n\]

The correlation coefficient is a measure of the strength of the linear relationship between two variables. The formula is given by: 

\[ r_{XY}= \frac{ cov_{XY} }{(\sum_{i=1}^{n}\sqrt{X-X_{i}} )(\sum_{i=1}^{n}\sqrt{Y-Y_{i}})} \]
\\
We used the correlation coefficient to detect the strength of linear relation between the variables.
\subsubsection*{ANOVA Table}Analysis of Variance or ANOVA table has been found out for different pairs and the explanatory power of the regressions has been studied.
\subsubsection*{Augmented Dickey Fuller Test:}  We performed the Augmented Dickey Fuller Test[39] for all three time series viz. NSE-Nifty index, Gold Prices and Crude Oil Prices. It helped us to check whether they have unit root or not[39]. Test for the presence of unit root is necessary so as to check whether the given data  is stationary or not[40]. It is necessary as it eases the data analysis of the data under consideration. We then performed the same test on their first order difference.\\ The Augmented Dickey Fuller Test is given below: \[ \Delta y_{t} = \alpha + \beta *t + \gamma y_{t-1} + \delta_{1} \Delta * y_{t-1} + \cdots \delta_{p-1}* \Delta *y_{t-p+1} + \epsilon_{t} \] \\The unit root test is then carried out for the null hypothesis that \(\gamma =0 \) against the alternative hypothesis that \(\gamma < 0 \). Then the value of the test statistic is calculated and if the test statistic is less than the larger critical value, then the null hypothesis is rejected and no unit root is present.
\subsection{Phillips-Perron Tests} We then use the Phillips-Perron Test to detect the presence of any \textit{Unit Root} present in the data. The only difference between Augmented Dickey Fuller Test and Phillips-Pearson(PP) test is that the PP test makes a non-parametric correction to the t-statistic[56].
\subsection{KPSS} We then performed the KPSS test[57] in order to detect the presence of any \textit{Unit root} in the three time series data.
\subsubsection*{Johansen Co-integration Tests} We then performed the Johansen Co-integration Integration [48]on the time-series of NSE-Nifty index, Gold and Crude prices. It is basically a linear combination of the I(1) time series $ (X_{1},X_{2},X_{3},....,X_{k})  $ to get a new time series  \[Y=b_{1}X_{1}+b_{2}X_{2}+....+b_{n}X_{k} \], which is I(0),  i.e. integrated of order zero (0). 
\subsubsection*{Fourier Transform} Fourier Transform of a function f(x), denoted by  I[f(x)] , is defined as \\
\[ I[f(x)]=F(s) = \int_{x_{1}}^{x_{2}} f(x) e^{isx} dx \] \\
We performed the Fourier Transform in order to detect if any frequency \textit{s} is present in the data? Where \textit{s} is the independent variable in this case.
\subsubsection{Wavelet Analysis} We perform the Granger Causality test[38] on the data obtained by operating on them the \textit{Discrete Wavelet Transform}. That is, basically we perform Granger Causality Test on different frequencies.
\paragraph{Discrete Wavelet Transform}We used the Haar Wavelet Transform[41] to decompose the signal into different frequencies. Haar Scaling Function is the simple unit-width, simple unit-height pulse function \(\phi (t)\) and \(\phi(2t)\) can be constructed from \(\phi (t)\) by \[\phi(t)= \phi(2t) + \phi(2t-1)\] \textit{Haar Wavelet  Transform} does not do the sampling and interpolation, thereby keeping the information intact. Here we represent the data as a linear combination of wavelet coefficients and scale coefficients.\\
As a matter of fact, if we want a construction from fine scale to coarser scale, the expansion coefficients at a lower level can be derieved from a higher scale by the following relation: \\
\[  c_{j}= \sum_{m} h(m-2k)c_{j+1}(m)\]  where j,k \(\in \mathbf{Z}, \mathbf{Z} \) is the set of integers.
\subparagraph{} And if we want a reconstruction from coarse scale to fine scale, the expansion coefficients of higher scale can be derieved from lower scale by the following relation:  \\
\subparagraph{}
\[c_{j+1}(k)=\sum_{m} c_{j}(m)h(k-2m) + \sum_{m} d_{j} (m) h_{1}(k-2m) \] \\ where, \(c_{i}\) represents the \(i^{th}\) scaling coefficients; \(d_{j}\) is the \(j^{th}\) Wavelet coefficients, h(x) being the filter.
Discrete wavelet transform decomposes the data by separating the high frequency and low frequency. \\
It has got the inverse relationship with scale. By increasing the frequencies we get lower scales. It is somewhat similar to Heisenberg's uncertainity principle, we cannot measure position and the velocity of an object exactly, simultaneously.
\subparagraph{Granger Causality}  In 1969, Granger Causality was proposed by C.W.J. Granger [38] which basically says, a time-series X is said to Granger cause Y, if the present X values can predict future Y values. We used the Vector Auto Regression (VAR) model to find the causality relation among the wavelet coefficients of Oil, Gold and Stock Index. Vector Autoregressive  Model is performed in a multi-variate framework, with k-dimensional multi-variate time-series. A VAR model is represented as follows: \\
\[ X(t)= a_{1}X(t-1) + a_{2}X(t-2) + a_{3}X(t-3) + ... + a_{k}X(t-k) + \epsilon (t) \] \\
where \(\epsilon(t)\) is the white noise. Our job is to find at least one \(a_{i}\), so that we can declare that the time series \(X_{i}\) can Granger cause \(X_{j}\). This marks the end of Data Analysis as far as time domain and frequency domain are concerned. Now we shall move towards Data Analysis in joint Time-Frequency domain. 

\subparagraph{Continuous Wavelet Transform:} In the joint Time-Frequency domain, we study the relationship between Gold Price \& NSE-Nifty and Crude Oil Price \& NSE-Nifty Index for times and different time-period(or frequencies).We find out the Wavelet Coherence between Oil \& NSE-Nifty and Gold \& NSE-Nifty. Then the multiple wavelet coherence(being defined later) has been calculated and subsequently it has been compared with wavelet coherence(being defined later) of the pairs. But to do so, by virtue of Heisenberg's uncertainity principle, we lose the information of either the frequency part or the time part. \\
A wavelet can be defined [50,51,52,53] as a real valued function \( $$ \psi (.) :\Re \rightarrow \Re $$ \) such that \\
\( 1. \int_{\Re} \psi (t) dt = 0  \) \\
\(2. \int_{-\infty}^{\infty}|\psi(t)|^{2} dt=1 \)
We then choose a reference wavelet known as mother wavelet. We have
\[\psi_{a,b} (t) = \frac{1}{\sqrt{a}} \psi (\frac{t-b}{a}) \] \\ provided \(a \neq 0\) and \textit{b} are real constants. Here, we are having \textit{a} as the scaling parametre and \textit{b} as the translation parameter. Unlike Fourier analysis, we have a variety of mother Wavelets, which we can use based on our requirements.\\ Thus, we define the Continuous Wavelet Transform as 
\[ W_{\psi}[f](a,b) = \int_{-\infty}^{\infty} f(t) \overline{\psi(\frac{t-u}{s})} dt \] \\ , where \( \overline{\psi} \) is the complex conjugate of \(\psi\) .From the square of the amplitude \(|W_{x}|^{2}\), known as the Wavelet Power Spectrum, we can determine the variance of the time-series under consideration for various frequencies, by virtue of Wavelet Power Spectrum; where large power indicates higher variances between the two time series and vice versa.
The Continuous Wavelet Transform should satisfy the following condition: \[C_{\psi} = \int_{-\infty}^{\infty} \frac{|\psi(\omega)|^{2}}{|\omega|} d\omega <\infty \] 
\paragraph{Wavelet Coherence} This is basically the correlation counterpart in time-frequency domain of the correlation in time domain. We have the Wavelet Transform of two Time-series x(t) and y(t) as \(W_{n}^{xy}= W_{n} ^{x} W_{n}^{y*}\), \(W_{n}^{y*}\)  represents the complex conjugate of the wavelet transform of the time series y(t). Then we get the cross-wavelet power spectrum as \(|W_{n}^{xy}|\). This will give us the variance between the two time series at different frequencies. The mathematical formula for the wavelet coherence is given below: \[R(x,y)=\frac{|S(s^{-1}W_{n}^{xy})|}{{S(s^{-1}|W_{n}^{x}|)^{\frac{1}{2}}}{S(s^{-1}|W_{n}^{y}|)^{\frac{1}{2}}}}\]  S being the smoothing process.[54,55]
\paragraph{Multiple Wavelet Coherence} Multiple Wavelet Coherence is just the counterpart of multiple correlation. It is basically done for time-frequency domain and is useful to detect wavelet coherence of a group of independent time series on a dependent one. For three time series x(t), y(t) and z(t) we get;
\[R(x,y)=\frac{|S(s^{-1}W_{n}^{xy})|}{{S(s^{-1}|W_{n}^{x}|)^{\frac{1}{2}}}{S(s^{-1}|W_{n}^{y}|)^{\frac{1}{2}}}}\]
\[R^{2}(y,x) = R(y,x).R(y,x)^{*} \]
\[R(x,z)=\frac{|S(s^{-1}W_{n}^{xz})|}{{S(s^{-1}|W_{n}^{x}|)^{\frac{1}{2}}}{S(s^{-1}|W_{n}^{z}|)^{\frac{1}{2}}}}\]
\[R^{2}(z,x) = R(z,x).R(z,x)^{*} \]
\[R(z,y)=\frac{|S(s^{-1}W_{n}^{zy})|}{{S(s^{-1}|W_{n}^{z}|)^{\frac{1}{2}}}{S(s^{-1}|W_{n}^{y}|)^{\frac{1}{2}}}}\]
\[R^{2}(y,z) = R(y,z).R(y,z)^{*} \] \\ We then calculate the Multiple Wavelet Coherence as follows: \\ \[RM^{2}(z,x,y) = \frac{R^{2}(z,y)+R^{2}(z,x)- 2Re[R(z,y).{R(z,x)}^{*}.{R(x,y)}^{*}]}{1- R^{2}(x,y)}\] \\ This will give the amount of the contribution of the independent time series x(t), y(t) on the dependent time series z(t) at a specific time and frequency.
\section{Programming, Results and Discussions} The codes for this research have been written partly in R Programming Language and partly in Python.\\ To specifically study the relation between the three time series we first studied the correlation between Crude \& NSE-Nifty and then Gold \& NSE-Nifty for three different periods. The specific time has been chosen based on our visual inspection. Then we studied the ANOVA table followed by ADF tests. These were part of our time domain tests. As far as frequency domain is concerned, we studied the Fourier Transform result to detect any hidden frequency. Subsequently, we decomposed the original time series and studied by virtue of the Granger Causality for various frequencies. 

\subsection{Correlation} By visual inspection, we find three periods where the participating time series will be either positively or negatively correlated. \\ The first period is chosen from  0 week to 200 week, the next period is chosen from 201 week to 700 week and the last period is chosen from 700 week to 1200 week. The results are shown below:\\
\begin{center}
	\begin{tabular}{|c|c|c|c|}
		\hline
		$ Period$ & $ r_{NG} $ & $ r_{GO} $ & $ r_{NO} $ \\ [0.5ex]
		\hline
		0-200 weeks & -0.22601402 & 0.2157788 & 0.33104342\\
		\hline
		200-700 weeks & 0.8840319 & 0.83914254 & 0.8587848\\
		\hline
		700-1200 weeks & -0.55126758 & 0.52345134 & -0.25074974\\ [1ex]
		\hline
	\end{tabular}
\end{center} 
 The weekly data time series from  5  November 1995 to 8 August 2018 has been shown in the Fig. 1.: \\
\begin{figure}
	\includegraphics[width=\linewidth]{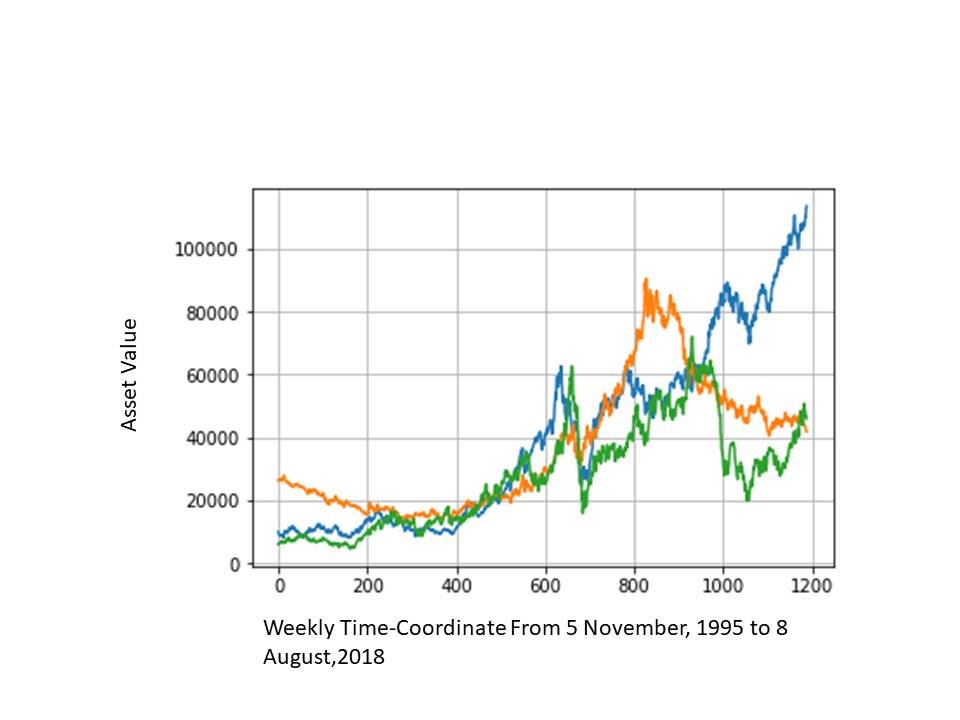}
	\caption{The graph depicting the asset value movements during the period Nov 1995 to Aug 2018.The asset values of Crude, Gold and NSE-Nifty are  expressed in India National Rupees. The color codes are:  Green ( Crude), Orange ( Gold) and Blue ( NSE-Nifty).}
	\label{fig:Graph}
\end{figure}
\\ As we can see from the graph that till 200 week Crude and NSE-Nifty were a bit correlated but Gold was not correlated with them. From 200 - 700 weeks the three time series were correlated and after 700 week, it has got messed up. All these are evident by simple inspection. But it should be quantified. Also we can see that there is a frequency mismatch. In the next ANOVA table, we can see the regression part:\\

\begin{figure}
	\includegraphics[width=\linewidth]{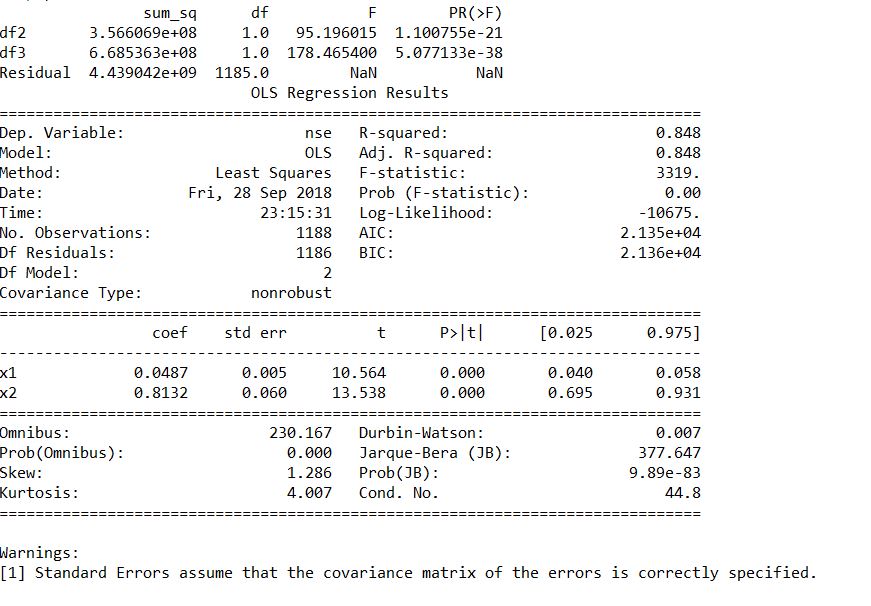}
	\caption{The ANOVA table showing the power of regression for the three time series}
	\label{fig:Graph}
\end{figure}
\subsection{ANOVA Table} From the ANOVA table (Fig. 2) we can see that the coefficients are statistically significant. The F-statistic is also showing that the coefficients are statistically significant at 5\%  and 2.5\% . Here we can see the Adj. R-squared is 0.848 which means that 84.8 percent of the variations of the dependent variable is explained by the independent variable. But the Durbin-Watson test shows that the error terms are positively correlated. So there may be any non-linear relationships.\subsection{Augmented Dickey Fuller Test}
We, then carried out the ADF test to find out the order of integration for the three time series, i.e. Crude, Gold and NSE-Nifty. \\
Trend: Constant
\begin{center}
	\begin{tabular}{|c|c|c|c|}
		\hline
		$ Time Series$ & $ ADF Statistics $ & $ p.value $ & $lags$ \\   
		\hline
		Oil &  -1.908 & 0.329 & 10 \\
		\hline
		$ \Delta Oil $ & -9.327 & 0.000000 & 9 \\
		\hline
		Gold &  -0.985 &  0.759 & 21 \\ 
		\hline
		$\Delta Gold$ & -6.277 & 0.000000 & 23\\
		\hline
		NSE-Nifty &  1.071 & 0.995 & 3 \\
		\hline
		$ \Delta NSE-Nifty $ & -19.773 & 0.000000 & 2\\
		\hline
	\end{tabular}
\end{center}

The critical value for different levels of significance for this test is: \\ 
\begin{center}
	\begin{tabular}{|c|c|}
		\hline
		$ Level of Significance $ & $ Critical Values $     \\ 
		\hline
		10 \% &  $ -2.57 $  \\
		\hline
		5 \%  & $ -2.86 $  \\
		\hline
		1 \% & $ -3.44 $ \\
		\hline
		
	\end{tabular}
\end{center}
\newpage
Trend: No Trend
\begin{center}
	\begin{tabular}{|c|c|c|c|}
		\hline
		$ Time Series$ & $ ADF Statistics $ & $ p.value $ & $lags$  \\ 
		\hline
		Oil &  -0.409 & 0.533 & 10 \\
		\hline
		$ \Delta Oil $ & -9.307 & 0.000000 & 9 \\
		\hline
		Gold &  -0.109 &  0.647 & 21 \\ 
		\hline
		$\Delta Gold$ & -6.265 & 0.000000 & 23\\
		\hline
		NSE-Nifty &  2.620 & 0.999 & 3 \\
		\hline
		$ \Delta NSE-Nifty $ & -22.494 & 0.000000 & 1\\
		\hline
	\end{tabular}
\end{center}

The critical value for different levels of significance for this test is: \\ 
\begin{center}
	\begin{tabular}{|c|c|}
		\hline
		$ Level of Significance $ & $ Critical Values $     \\ 
		\hline
		10 \% &  $ -1.62 $  \\
		\hline
		5 \%  & $ -1.94 $  \\
		\hline
		1 \% & $ -2.57 $ \\
		\hline
		
	\end{tabular}
\end{center}
\newpage
Trend: Constant and Linear Trend
\begin{center}
	\begin{tabular}{|c|c|c|c|}
		\hline
		$ Time Series$ & $ ADF Statistics $ & $ p.value $ & $lags$  \\ 
		\hline
		Oil &  -3.180 & 0.088 & 10 \\
		\hline
		$ \Delta Oil $ & -9.324 & 0.000000 & 9 \\
		\hline
		Gold &  -1.163 & 0.918 & 21 \\ 
		\hline
		$\Delta Gold$ & -6.277 & 0.000000 & 23\\
		\hline
		NSE-Nifty &  -1.714 & 0.745 & 3 \\
		\hline
		$ \Delta NSE-Nifty $ & -19.865 & 0.000000 & 2\\
		\hline
	\end{tabular}
\end{center}

The critical value for different levels of significance for this test is: \\ 
\begin{center}
	\begin{tabular}{|c|c|}
		\hline
		$ Level of Significance $ & $ Critical Values $     \\ 
		\hline
		10 \% &  $ -3.13 $  \\
		\hline
		5 \%  & $ -3.41 $  \\
		\hline
		1 \% & $ -3.97 $ \\
		\hline
		
	\end{tabular}
\end{center}

 So clearly we can see that it is a I(1) series i.e. it is integrated of order one (1). 
\subsection{Phillips-Perron Test}
We, then carried out the Phillips-Pearson test to find out the order of integration for the three time series, i.e. Crude, Gold and NSE-Nifty.
\newpage
Trend: Constant
\begin{center}
	\begin{tabular}{|c|c|c|c|}
		\hline
		$ Time Series$ & $ ADF Statistics $ & $ p.value $ & $lags$  \\ 
		\hline
		Oil &  -1.745 & 0.408 & 23 \\
		\hline
		$ \Delta Oil $ & -33.751 & 0.000000 & 23 \\
		\hline
		Gold &  -0.907 &  0.786 & 23 \\ 
		\hline
		$\Delta Gold$ & -34.238 & 0.000000 & 23\\
		\hline
		NSE-Nifty &  1.121 & 0.995 & 23 \\
		\hline
		$ \Delta NSE-Nifty $ & -33.306 & 0.000000 & 23\\
		\hline
	\end{tabular}
\end{center}

The critical value for different levels of significance for this test is: \\ 
\begin{center}
	\begin{tabular}{|c|c|}
		\hline
		$ Level of Significance $ & $ Critical Values $     \\ 
		\hline
		10 \% &  $ -2.57 $  \\
		\hline
		5 \%  & $ -2.86 $  \\
		\hline
		1 \% & $ -3.44 $ \\
		\hline
		
	\end{tabular}
\end{center}
\newpage 
Trend: No Trend
\begin{center}
	\begin{tabular}{|c|c|c|c|}
		\hline
		$ Time Series$ & $ ADF Statistics $ & $ p.value $ & $lags$  \\ 
		\hline
		Oil &  -0.257 &  0.593 & 23 \\
		\hline
		$ \Delta Oil $ & -33.770 & 0.000000 & 23 \\
		\hline
		Gold &  -0.032 &  0.674 & 23 \\ 
		\hline
		$\Delta Gold$ & -34.243 & 0.000000 & 23\\
		\hline
		NSE-Nifty &  2.667 & 0.999 & 23 \\
		\hline
		$ \Delta NSE-Nifty $ & -33.276 & 0.000000 & 23\\
		\hline
	\end{tabular}
\end{center}

The critical value for different levels of significance for this test is: \\ 
\begin{center}
	\begin{tabular}{|c|c|}
		\hline
		$ Level of Significance $ & $ Critical Values $     \\ 
		\hline
		10 \% &  $ -1.62 $  \\
		\hline
		5 \%  & $ -1.94 $  \\
		\hline
		1 \% & $ -2.57 $ \\
		\hline
		
	\end{tabular}
\end{center}
\newpage
Trend: Constant and Linear Trend
\begin{center}
	\begin{tabular}{|c|c|c|c|}
		\hline
		$ Time Series$ & $ ADF Statistics $ & $ p.value $ & $lags$  \\ 
		\hline
		Oil &  -2.900 & 0.162 & 23 \\
		\hline
		$ \Delta Oil $ & -33.738 & 0.000000 & 23 \\
		\hline
		Gold &  -1.106 & 0.928 & 23 \\ 
		\hline
		$\Delta Gold$ & -34.224 & 0.000000 & 23\\
		\hline
		NSE-Nifty &  -1.827 & 0.692 & 23 \\
		\hline
		$ \Delta NSE-Nifty $ & -33.348 & 0.000000 & 23\\
		\hline
	\end{tabular}
\end{center}

The critical value for different levels of significance for this test is: \\ 
\begin{center}
	\begin{tabular}{|c|c|}
		\hline
		$ Level of Significance $ & $ Critical Values $     \\ 
		\hline
		10 \% &  $ -3.13 $  \\
		\hline
		5 \%  & $ -3.41 $  \\
		\hline
		1 \% & $ -3.97 $ \\
		\hline
		
	\end{tabular}
\end{center}

 So clearly we can see that it is a I(1) series i.e. it is integrated of order one (1). 
\subsection{KPSS}
We, then carried out the Phillips-Pearson test to find out the order of integration for the three time series, i.e. Crude, Gold and NSE-Nifty.
\newpage
Trend: Constant
\begin{center}
	\begin{tabular}{|c|c|c|c|}
		\hline
		$ Time Series$ & $ ADF Statistics $ & $ p.value $ & $lags$  \\ 
		\hline
		Oil &  3.783 & 0.000000 & 23 \\
		\hline
		$ \Delta Oil $ & 0.033 & 0.965 & 23 \\
		\hline
		Gold &  3.456 &  0.000 & 23 \\ 
		\hline
		$\Delta Gold$ & 0.255 & 0.182 & 23\\
		\hline
		NSE-Nifty & 4.647 & 0.000 & 23 \\
		\hline
		$ \Delta NSE-Nifty $ & 0.308 & 0.129 & 23\\
		\hline
	\end{tabular}
\end{center}

The critical value for different levels of significance for this test is: \\ 
\begin{center}
	\begin{tabular}{|c|c|}
		\hline
		$ Level of Significance $ & $ Critical Values $     \\ 
		\hline
		10 \% &  $ 0.35 $  \\
		\hline
		5 \%  & $ 0.46 $  \\
		\hline
		1 \% & $ 0.74 $ \\
		\hline
		
	\end{tabular}
\end{center}
\newpage
Trend: Constant and Linear Trend
\begin{center}
	\begin{tabular}{|c|c|c|c|}
		\hline
		$ Time Series$ & $ ADF Statistics $ & $ p.value $ & $lags$  \\ 
		\hline
		Oil &   0.394 & 0.000 & 23 \\
		\hline
		$ \Delta Oil $ &  0.031 & 0.856 & 23 \\
		\hline
		Gold &  0.508 & 0.000 & 23 \\ 
		\hline
		$\Delta Gold$ &  0.251 & 0.005 & 23\\
		\hline
		NSE-Nifty &  0.715 & 0.000 & 23 \\
		\hline
		$ \Delta NSE-Nifty $ & 0.029 & 0.883 & 23\\
		\hline
	\end{tabular}
\end{center}
 \(\Delta\) being the first order differences \\
The critical value for different levels of significance for this test is: \\ 
\begin{center}
	\begin{tabular}{|c|c|}
		\hline
		$ Level of Significance $ & $ Critical Values $     \\ 
		\hline
		10 \% & $ 0.12 $  \\
		\hline
		5 \%  & $ 0.15 $  \\
		\hline
		1 \% & $ 0.22 $ \\
		\hline
		
	\end{tabular}
\end{center}

 So clearly we can see that it is a I(1) series i.e. it is integrated of order one (1). \\ \textbf{ \( \Delta\) being the first order difference }
\subsection{Johansen Cointegration Test} We now go for the Johansen cointegration test which is done by the help of R \textit{urca} library. The results are shown below:\\
Test type: trace statistic , with linear trend \\

Eigenvalues (lambda):
\\
\begin{center}
	\begin{tabular}{|c|c|c|}
		\hline
		$  0.0088648918 $ & $0.0013532634 $ & $0.0006105299$     \\ 
		\hline
		
	\end{tabular}
	
\end{center}
  Values of test statistic and critical values of test:

\begin{center}
	\begin{tabular}{|c|c|c|c|c|}
		\hline
		$  Hypothesis $ & $ test statistics $ & $ 10 pct$ & $5 pct$ & $1 pct$ \\
		\hline
		$r<=2 $ & $0.72$ & $ 6.50 $ & $8.18$ & $11.65$\\
		$r<=1 $ & $2.33 $ & $15.66$ & $17.95$ & $23.52$ \\
		$r=0 $ & $12.89$ & $28.71$ & $31.52$ & $37.22$ \\
		\hline
		
	\end{tabular}
	
\end{center}
  Eigenvectors, normalised to first column:
(These are the cointegration relations)
\\
\begin{center}
	\begin{tabular}{|c|c|c|c|c|}
		\hline
		$  Hypothesis $ & $ a.12 $ & $ b.12 $ & $c.12$     \\
		$a.12 $ & $1.00000000$ & $ 1.00000000 $ & $ 1.00000000 $\\
		$b.12 $ & $-0.06039987 $ & $-0.2392422$ & $0.2546972$  \\
		$c.12 $ & $-0.06972624$ & $-2.2453992$ & $-2.0694282$  \\
		\hline
		
	\end{tabular}
	
\end{center}
 Taking the Eigenvectors, we can formulate an equation, \(s=1*Oil - 0.06039987*Gold - 0.06972624*NSE-Nifty\) and plot it graphically in Fig. 3 to see whether it is Stationary or not. Moreover from the table above we can very well understand that the time-series are not co-integrated.\\
\begin{figure}
	\includegraphics[width=\linewidth]{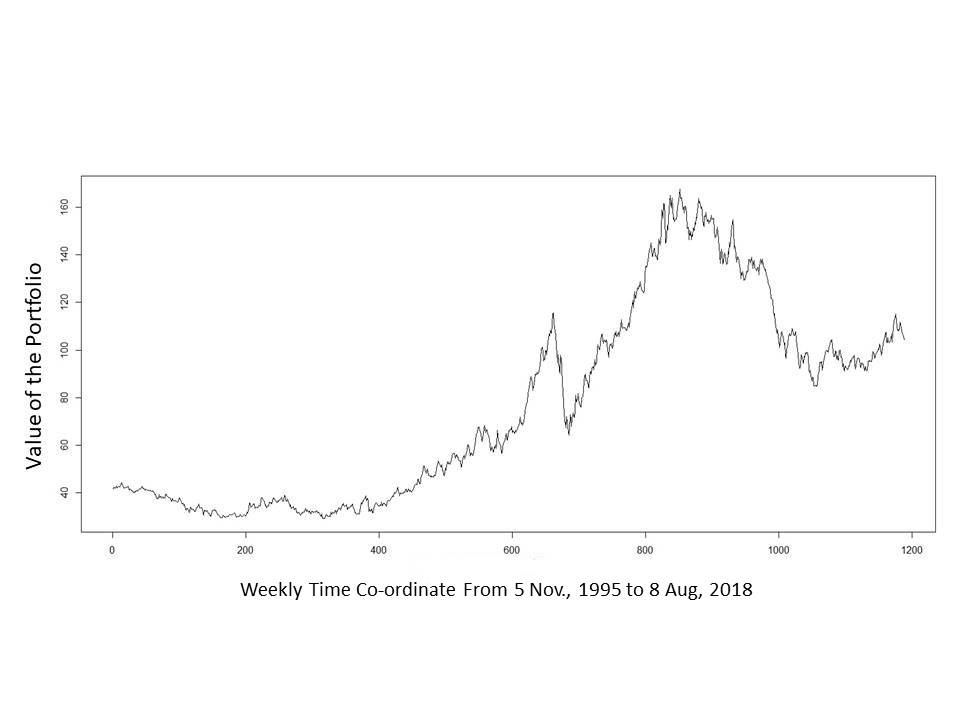}
	\caption{The plotting of the cointegrated portfolio. The portfolio  has been calculated using the relation \(s=1*Oil - 0.06039987*Gold - 0.06972624*NSE-Nifty\) , the details of which are discussed in Section 3.4}
	\label{fig:Graph}
\end{figure}
\newpage
\subsection{Discrete Wavelet Transform}
Signif. codes:  0 ‘***’ 0.001 ‘**’ 0.01 ‘*’ 0.05 ‘.’ 0.1 ‘ ’ 1\\
\\ Scale 1 (2-4 weeks)\\

\begin{center}
	\begin{tabular}{|c|c|c|c|}
		\hline
		$  Dependent Variable $ & $ Independent variable $ & $ F-stat $ & $Prob.$     \\
		$Crude $ & $ NSE-Nifty $ & $ 4.1512 $ & $ 0.006325 ** $\\
		$NSE-Nifty $ & $Crude $ & $0.9627$ & $0.41$  \\
		$Gold $ & $NSE-Nifty$ & $0.4798$ & $0.6964$  \\
		$NSE-Nifty $ & $Gold$ & $1.7937$ & $0.1472$  \\		
		\hline
		
	\end{tabular}
	
\end{center} Scale 2 (4-8 weeks)\\

\begin{center}
	\begin{tabular}{|c|c|c|c|}
		\hline
		$  Dependent Variable $ & $ Independent variable $ & $ F-stat $ & $Prob.$     \\
		$Crude $ & $ NSE-Nifty $ & $ 2.2017 $ & $ 0.08804 . $\\
		$NSE-Nifty $ & $Crude $ & $1.3195$ & $0.2682$  \\
		$Gold $ & $NSE-Nifty$ & $1.9166$ & $0.127$  \\
		$NSE-Nifty $ & $Gold$ & $2.0308$ & $0.1097$  \\		
		\hline
		
	\end{tabular}
	
\end{center}
 Scale 3 (8-16 weeks)\\

\begin{center}
	\begin{tabular}{|c|c|c|c|}
		\hline
		$  Dependent Variable $ & $ Independent variable $ & $ F-stat $ & $Prob.$     \\
		$Crude $ & $ NSE-Nifty $ & $ 1.2056 $ & $ 0.3101 $\\
		$NSE-Nifty $ & $Crude $ & $0.8665$ & $ 0.4602$  \\
		$Gold $ & $NSE-Nifty$ & $0.9789$ & $0.4047$  \\
		$NSE-Nifty $ & $Gold$ & $0.5528$ & $0.6471$  \\		
		\hline
		
	\end{tabular}
	
\end{center}
\newpage
 Scale 4 (16-32 weeks)\\

\begin{center}
	\begin{tabular}{|c|c|c|c|}
		\hline
		$  Dependent Variable $ & $ Independent variable $ & $ F-stat $ & $Prob.$     \\
		$Crude $ & $ NSE-Nifty $ & $ 2.0094 $ & $ 0.1215 $\\
		$NSE-Nifty $ & $Crude $ & $3.1395$ & $ 0.03129 *$  \\
		$Gold $ & $NSE-Nifty$ & $0.996$ & $0.4005$  \\
		$NSE-Nifty $ & $Gold$ & $0.9025$ & $0.4449$  \\		
		\hline
		
	\end{tabular}
	
\end{center}
 Scale 5 (32-64 weeks)\\
\begin{center}
	\begin{tabular}{|c|c|c|c|}
		\hline
		$  Dependent Variable $ & $ Independent variable $ & $ F-stat $ & $Prob.$     \\
		$Crude $ & $ NSE-Nifty $ & $ 0.9623 $ & $ 0.4248 $\\
		$NSE-Nifty $ & $Crude $ & $0.306$ & $ 0.8208$  \\
		$Gold $ & $NSE-Nifty$ & $1.6423$ & $0.203$  \\
		$NSE-Nifty $ & $Gold$ & $ 0.5822$ & $0.6318$  \\		
		\hline
		
	\end{tabular}
	
\end{center}
 Scale 6 (64-128 weeks)\\
\begin{center}
	\begin{tabular}{|c|c|c|c|}
		\hline
		$  Dependent Variable $ & $ Independent variable $ & $ F-stat $ & $Prob.$     \\
		$Crude $ & $ NSE-Nifty $ & $ 1.8376 $ & $ 0.2185 $\\
		$NSE-Nifty $ & $Crude $ & $1.5133$ & $ 0.2838$  \\
		$Gold $ & $NSE-Nifty$ & $1.3021$ & $ 0.3389$  \\
		$NSE-Nifty $ & $Gold$ & $ 0.3033$ & $0.8224$  \\		
		\hline
		
	\end{tabular}
	
\end{center}
\newpage
 Scale 7 (128-256 weeks)\\
\begin{center}
	\begin{tabular}{|c|c|c|c|}
		\hline
		$  Dependent Variable $ & $ Independent variable $ & $ F-stat $ & $Prob.$     \\
		$Crude $ & $ NSE-Nifty $ & $ 1.8376 $ & $ 0.2185 $\\
		$NSE-Nifty $ & $Crude $ & $1.5133$ & $ 0.2838$  \\
		$Gold $ & $NSE-Nifty$ & $1.3021$ & $ 0.3389$  \\
		$NSE-Nifty $ & $Gold$ & $ 0.3033$ & $0.8224$  \\		
		\hline
		
	\end{tabular}
	
\end{center}
\paragraph{} With the help of Haar a trous wavelet transform, we decompose the original time-series into different scales. There are seven scales all together, being decomposed. We then study the Granger Causalities for the two different pairs viz. NSE-Nifty \& Crude and  NSE-Nifty \& Gold. We have checked it by virtue of F-test. as seen from the tables, we could only find a relationship for the scale of 2-4 weeks; 4-8 weeks and 16-32 weeks. For the 2-4 weeks scale, we could see NSE-Nifty granger cause Crude at 1\% level of significance. For the scale of 4-8 weeks, NSE-Nifty granger cause Crude at 10\% level of significance. For the 16-32 weeks Crude Granger cause NSE-Nifty at 5\% level of significance. For other pairs there are no granger causal relations.\\ Unlike a number of nexuses as reported for the Chinese Market [19], as such here for the Indian markets, we are unable to find a large number of such nexuses.

\subsection{Fourier Transform} We performed the Fourier Transform (Fig. 4) for a range of frequencies in 0-500 \((week)^{-1}\) range inorder to detect any latent frequency present therein for the NSE-Nifty data. Unfortunately there is no frequency detected. As seen from the graph, it is highest for \(\omega = 0 week^{-1}\) (zero). As for explanation, 1 \((week)^{-1}\) is defined as one cycle per week. \\

\begin{figure}
	\includegraphics[width=20cm, height=15cm]{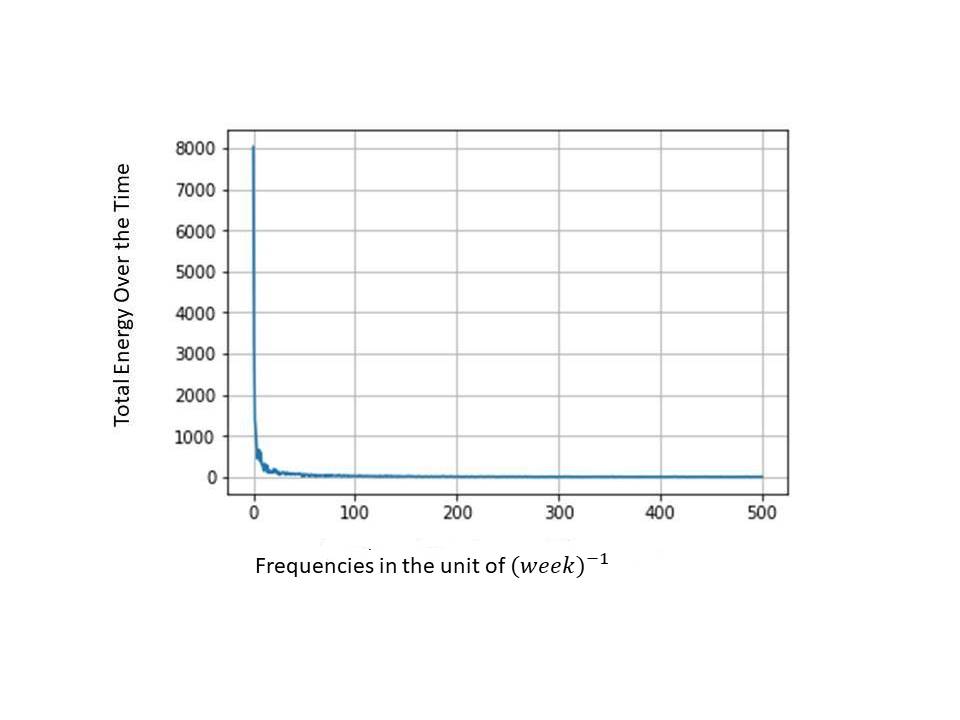}
	\caption{Fourier Transform for a range of frequencies in 0 to 500 (week) 1 range inorder to detect any latent frequency present therein for the NSE-Nifty data. (There is no frequency detected)}
	\label{fig:Graph}
\end{figure}
\subsection{Wavelet Coherence}
Wavelet Coherence works on a similar motive as that of a correlation method. We performed the \textit{Cross Wavelet Coherence Test} [62] for NSE-Nifty \& Gold, then NSE-Nifty \& Crude and Multiple Wavelet Coherence[62] for all the three. We have a heat map which indicates the amount of coherence between the participating time-series.\\Wavelet coherence graph of both NSE-Nifty \& Gold and NSE-Nifty \& Crude have been studied. Horizontal axis depicts the time in multiple of 50 weeks whereas vertical axis depicts the time-period. We  should note here, the unit of frequency used was "1/week" earlier. But now onwards for Wavelet Analysis we shall use the term \textit{week} ( always written in italics) to mean the frequency. \\ For the NSE-Nifty \& Gold nexuses (Fig. 5), we can see that in shorter time-periods there is no indication of coherence for all the time from November 1995 to August 2018. From 400 weeks to 800 weeks for the low frequency zones of 64-128 \textit{week}; 0 week to almost 800 weeks in the low frequency zones of 128-256 \textit{week}; and for initial few weeks of the data in all frequency levels, we get relatively better coherence for NSE-Nifty \& Gold. But still there are small patches of absence of high power levels in many inbetween cases. \\ Regarding crude oil (Fig. 6), there was initially no coherence for any of the low frequency zones. But from 2003, the wavelet coherence gradually increased, though it is not the best but still better.\\ It now appears from our studies that, there is coherence for low frequencies (64-128 \textit{week} and 128-256 \textit{week}), but no coherence for high frequencies (2-4 \textit{week}, 4-8 \textit{week}, 8-16 \textit{week}, 16-32 \textit{week} and 32-64 \textit{week}). For the initial few weeks in all frequencies; and in the 400-800 weeks in the low frequency zones of 64-128 \textit{week}; we get relatively better coherence as far as the NSE-Nifty \& Oil is concerned. For the NSE-Nifty, Gold and Oil multiple coherence test, we get a higher coherence in the initial few weeks in all the frequency zones. Later on, during 400-800 weeks in the frequency zones of 6-7 \textit{week} approximately, we get a good coherence. Thus for long term traders they can include gold and/or crude in their portfolio along with NSE-Nifty index in order to decrease the risk(volatility) of the portfolio for Indian Market. But for short term traders, it is advisable, not to include all the three in their portfolio.
\begin{figure}
	\includegraphics[width=\linewidth]{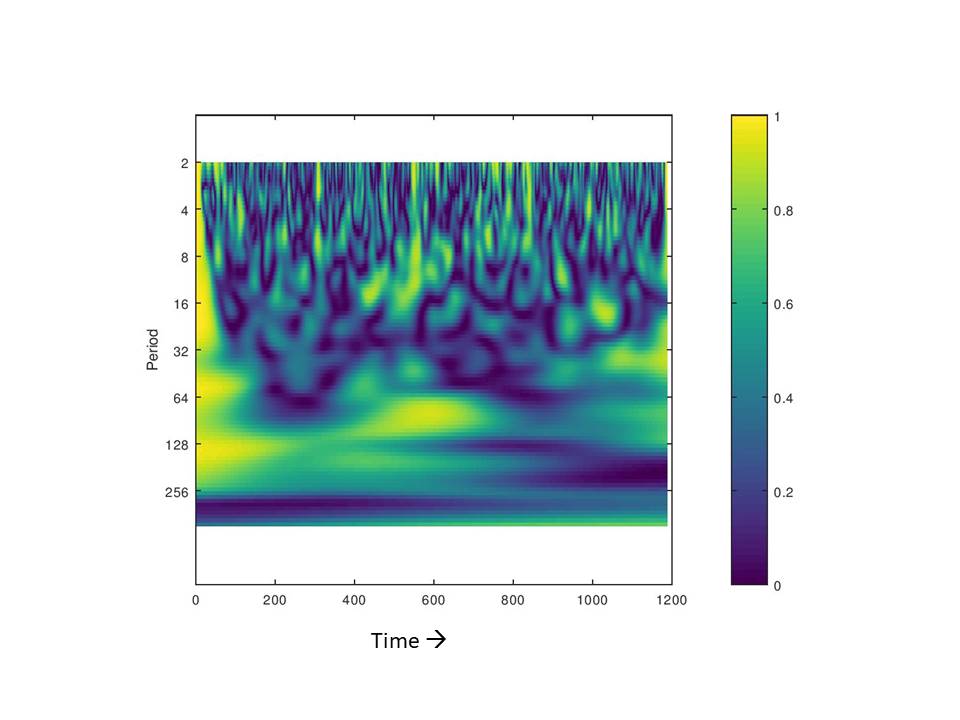}
	\caption{NSE-Nifty-Gold Wavelet Coherence. Here the Period (along vertical axis) is in week, and Time (along horizontal axis) is in week. The vertical scale on the right side is the normalized colour coded value of Power in the power spectrum. }
	\label{fig:Graph}
\end{figure}

\begin{figure}
	\includegraphics[width=\linewidth]{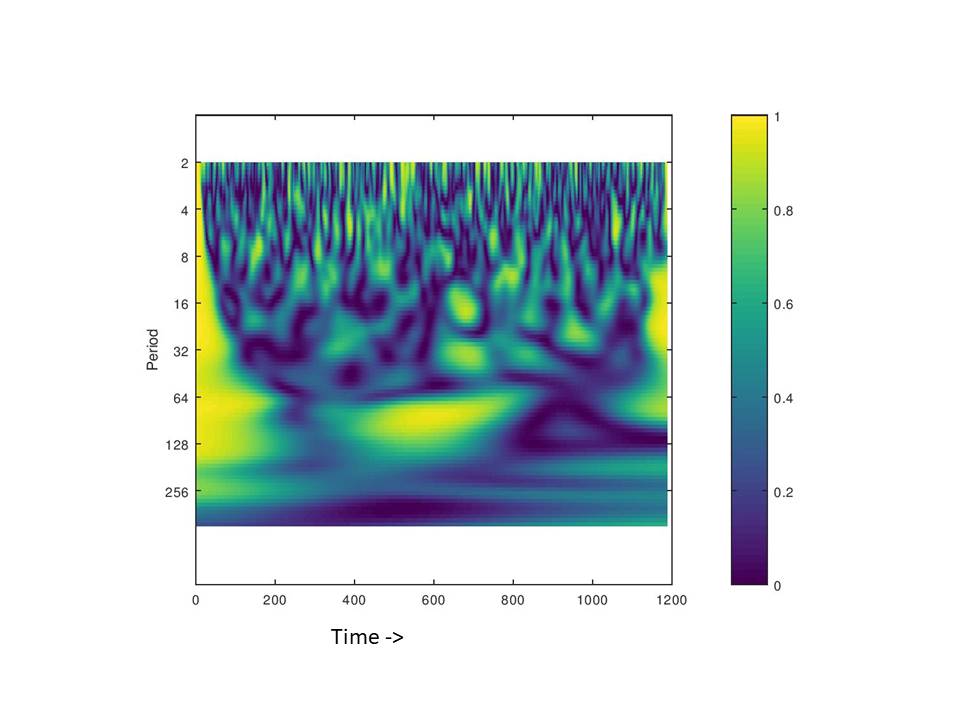}
	\caption{NSE-Nifty-Crude Wavelet Coherence. Here the Period (along vertical axis) is in week, and Time (along horizontal axis) is in week. The vertical scale on the right side is the normalized colour coded value of Power in the power spectrum.}
	\label{fig:Graph}
\end{figure}
\begin{figure}
	\includegraphics[width=\linewidth]{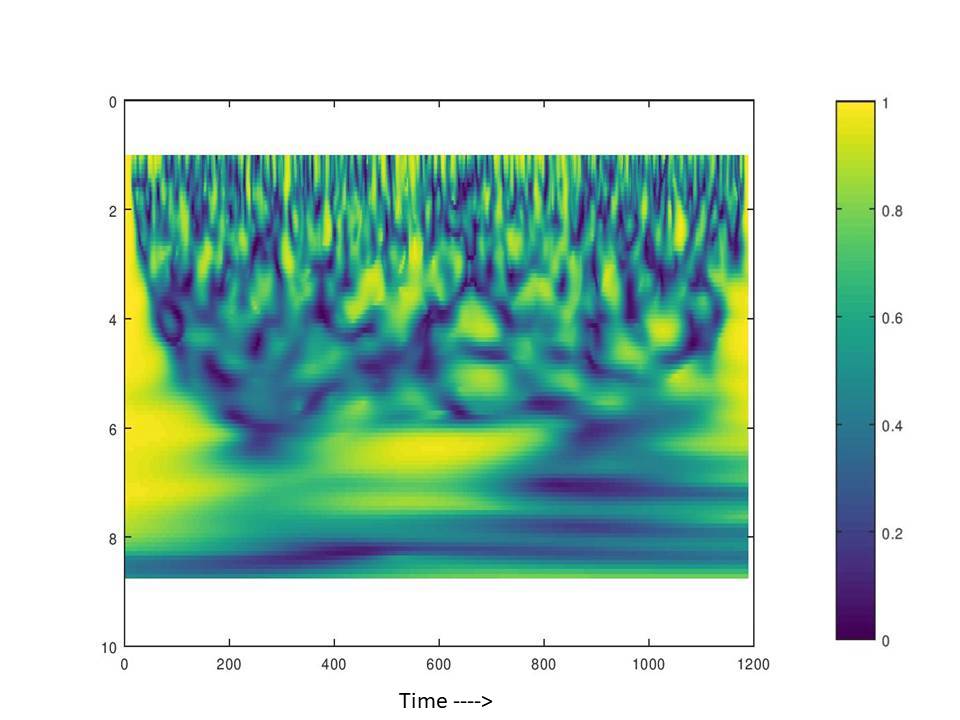}
	\caption{NSE-Nifty-Crude Multiple Wavelet Coherence. Here the Period (along vertical axis) is in week, and Time (along horizontal axis) is in week. The vertical scale on the right side is the normalized colour coded value of Power in the power spectrum.}
	\label{fig:Graph}
\end{figure}

\section{Conclusions}In this paper, we undertook a research based on the time domain, frequency domain and again on joint time-frequency domain for the three variables Crude, Gold and NSE-Nifty. The correlation tests indicate a good correlation on some particular time interval. The ANNOVA table indicates good regression but with a positive serial correlation. \textit{Cointegration tests} show that there is no relation among the three time-series data. We then perform the Fourier Transformation for frequency domain and get no corresponding frequencies for the time series. \textit{Granger Causality Tests} were performed. We did not find any economically viable causal relationships for any pairs with the help of \textit{Granger Causality Tests}. For joint time-frequency domain we perform the \textit{Continuous Wavelet Coherence} tests to determine the coherence between the two pair of time series. \\ With respect to the correlations, we find that the pairs; NSE-Nifty \& Gold and NSE-Nifty \& Crude are not correlated till first 200 weeks. But after \(200\) weeks till \(700\) weeks there is a very high correlation (more than 0.80 for both the pairs of time-series). This is basically a structure broken point. But after that, till August, 2018; we can not see any such correlation between the two pairs. Next we find an adjusted R-squared equal to 0.848, but the DW test shows a value of 0.7, which indicates positive serial correlation. Cointegration test shows no indication of any relationships between the three time-series data. Fourier Transform shows no presence of frequencies for the NSE-Nifty data.We find three Granger causal relations for different frequencies, but they are unidirectional. We then execute the \textit{Continuous Wavelet Transform} to detect the coherence test. A lot of coherence has been established for different time-zones at different frequencies. So we get to know, the otherwise relationships which had high serial-correlation, no Cointegration relationships and very few irrelevant Granger Causal relationships.\\
\begin{center}
	\begin{tabular}{|c|}
		\hline
		Summary of Analysis \\ [0.5ex]
		\hline
		\hline
		Correlation Tests:  As such no great Correlation found for the entire\\ period except from \(200^{th}\) week to \(700^{th}\) week \\ [0.5ex]
		\hline
		Regression Analysis: Error Terms positively Correlated\\ [1ex]
		\hline
		ADF, PP, KPSS Tests: Found the three data-sets are integrated of order one.\\ [1ex]  
		\hline
		Johansen Co-integration Test: No cointegration exists between the three data-sets\\ [1ex]
		\hline
		Fourier Test: No hidden frequency detected\\ [1ex] 
		\hline
		Granger Causality on Discrete Wavelet Transform: No economically viable\\ cause-effect relations detected\\ [1ex]
		\hline 
		Continuous Wavelet Transform: Some great coherence values are found corresponding \\ to some particular frequencies\\ [1ex]
		\hline 
	\end{tabular}
\end{center}
\newpage

\section*{References}
\begin{itemize}
	\item [1] K. Wang, CAN GOLD EFFECTIVELY HEDGE RISKS OF EXCHANGE RATE?, Journal Of Business Economics And Management. 14 (2013) 833-851. doi:10.3846/16111699.2012.670133.
	\item [2] M. Dooley, P. Isard, M. Taylor, Exchange rates, country-specific shocks, and gold, Applied Financial Economics. 5 (1995) 121-129. doi:10.1080/758522999.
	\item [3] S. Baker, R. van Tassel, Forecasting the price of gold: A fundamentalist approach, Atlantic Economic Journal. 13 (1985) 43-51. doi:10.1007/bf02304036.
	\item [4] B. Kolluri, Gold as a hedge against inflation: An empirical investigation, Quarterly Review Of Economics And Business. (1981) 13-24.
	Laurent 1994; Mahdavi, Zhou 1997; Moore 1990
	\item [5] L. Sjaastad, F. Scacciavillani, The price of gold and the exchange rate, Journal Of International Money And Finance. 15 (1996) 879-897. doi:10.1016/s0261-5606(96)00045-9.
	\item [6] F. Capie, T. Mills, G. Wood, Gold as a hedge against the dollar, Journal Of International Financial Markets, Institutions And Money. 15 (2005) 343-352. doi:10.1016/j.intfin.2004.07.002.
	\item [7] D. Baur, B. Lucey, Is Gold a Hedge or a Safe Haven? An Analysis of Stocks, Bonds and Gold, Financial Review. 45 (2010) 217-229. doi:10.1111/j.1540-6288.2010.00244.x.
	\item [8] T. Ischuk, D. Zhilkin, T. Aikina, Gold as a Tool for Hedging Financial Risks, IOP Conference Series: Earth And Environmental Science. 43 (2016) 012085. doi:10.1088/1755-1315/43/1/012085.
	\item [9] M. Souček, Crude oil, equity and gold futures open interest co-movements, Energy Economics. 40 (2013) 306-315. doi:10.1016/j.eneco.2013.07.010.
	\item [10] D. Baur, B. Lucey, Is Gold a Hedge or a Safe Haven? An Analysis of Stocks, Bonds and Gold, Financial Review. 45 (2010) 217-229. doi:10.1111/j.1540-6288.2010.00244.x.
	\item[11]M. AROURI, P. Foulquier and J. Fouquau, "Oil Prices and Stock Markets in Europe: A Sector Perspective", Recherches économiques de Louvain, vol. 77, no. 1, p. 5, 2011.
	\item [12] M. Arouri and C. Rault, "OIL PRICES AND STOCK MARKETS IN GCC COUNTRIES: EMPIRICAL EVIDENCE FROM PANEL ANALYSIS", International Journal of Finance and Economics, vol. 17, no. 3, pp. 242-253, 2011
	
	\item[13]D. Asteriou, A. Dimitras and A. Lendewig, "The Influence of Oil Prices on Stock Market Returns: Empirical Evidence from Oil Exporting and Oil Importing Countries", International Journal of Business and Management, vol. 8, no. 18, 2013
	\item[14]G. Caporale, F. Menla Ali, N. Spagnolo, Oil price uncertainty and sectoral stock returns in China: A time-varying approach, China Economic Review. 34 (2015) 311-321. doi:10.1016/j.chieco.2014.09.008.
	\item[15]S. Ghosh and K. Kanjilal, "Co-movement of international crude oil price and Indian stock market: Evidences from nonlinear cointegration tests", Energy Economics, vol. 53, pp. 111-117, 2016.
	\item[16]E. Bouri, A. Jain, P. Biswal, D. Roubaud, Cointegration and nonlinear causality amongst gold, oil, and the Indian stock market: Evidence from implied volatility indices, Resources Policy. 52 (2017) 201-206. doi:10.1016/j.resourpol.2017.03.003.
	\item[17]L. Ping, Z. Ziyi, Y. Tianna, Z. Qingchao, The relationship among China’s fuel oil spot, futures and stock markets, Finance Research Letters. 24 (2018) 151-162. doi:10.1016/j.frl.2017.09.001.
	\item[18] P. Bairagi and V. Rai, "Impact of changes in Oil Price on Indian Stock Market", 2014
	\item[19]S. Huang, H. An, X. Gao, X. Huang, Time–frequency featured co-movement between the stock and prices of crude oil and gold, Physica A: Statistical Mechanics And Its Applications. 444 (2016) 985-995. doi:10.1016/j.physa.2015.10.080.
	\item[20]L. Yang, S. Tian, W. Yang, M. Xu, S. Hamori, Dependence structures between Chinese stock markets and the international financial market: Evidence from a wavelet-based quantile regression approach, The North American Journal Of Economics And Finance. 45 (2018) 116-137. doi:10.1016/j.najef.2018.02.005.
	\item[21]A. Bhunia, COINTEGRATION AND CAUSAL RELATIONSHIP AMONG CRUDE PRICE, DOMESTIC GOLD PRICE AND FINANCIAL VARIABLESAN EVIDENCE OF BSE AND NSE, Journal Of Contemporary Issues In Business Research. 2 (2013).
	\item[22]A. Bhunia, Relationships between Commodity Market Indicators and Stock Market Index-an Evidence of India, Academy Of Contemporary Research Journal. (2013) 5.
	\item[23] S. Huang, H. An, X. Gao, X. Huang, Time–frequency featured co-movement between the stock and prices of crude oil and gold, Physica A: Statistical Mechanics And Its Applications. 444 (2016) 985-995. doi:10.1016/j.physa.2015.10.080.
	\item[24] C. Albulescu, D. Goyeau, A. Tiwari, Contagion and Dynamic Correlation of the Main European Stock Index Futures Markets: A Time-frequency Approach, Procedia Economics And Finance. 20 (2015) 19-27. doi:10.1016/s2212-5671(15)00042-8.
	\item[25] S. Saravanan, S. Mala, Stock Market Prediction System: A Wavelet based Approach, Applied Mathematics and Information Sciences. 12 (2018) 579-585. doi:10.18576/amis/120312.
	\item[26] R. Deora, D. Nguyen, Time-Scale Comovement Between The Indian And World Stock Markets, The Journal Of Applied Business Research. 29 (2013) 12.
	\item[27] R. Deora, D. Nguyen, Time-Scale Comovement Between The Indian And World Stock Markets, The Journal Of Applied Business Research. 29 (2013) 12.
	\item[28] Jothimani, D., Shankar, R., Yadav, S.S., (2015) Discrete Wavelet Transform-Based Prediction of Stock Index: A Study on National Stock Exchange Fifty Index, Journal of Financial Management and Analysis, 28 (2), 35-49. 
	\item[29] D. Das, M. Kannadhasan, K. Al-Yahyaee, S. Yoon, A wavelet analysis of co-movements in Asian gold markets, Physica A: Statistical Mechanics And Its Applications. 492 (2018) 192-206. doi:10.1016/j.physa.2017.09.061.
	\item[30] A. Kumar, B. Kamaiah, Returns and volatility spillover between Asian equity markets: A wavelet approach, Ekonomski Anali. 62 (2017) 63-83. doi:10.2298/eka1712063k.
	\item[31] A. Shah, M. Deo, Integration of the Indian Stock Market : at the angle of Time-Frequency, Journal Of Economic Integration. 31 (2016) 183-205. doi:10.11130/jei.2016.31.1.183.
	\item[32] D. Workman, Crude Oil Imports by Country, World's Top Exports. (2018). http://www.worldstopexports.com/crude-oil-imports-by-country/ (accessed 13 August 2018).
	\item[33] Forbes.Com. (2018). https://www.forbes.com/sites/greatspeculations/2017/10/11/germans-have-quietly-become-the-worlds-biggest-buyers-of-gold/ (accessed 13 August 2018).
	\item[34] Nifty 50 Historical Rates - Investing.com, Investing.Com. (2018). https://www.investing.com/indices/s-p-cnx-nifty-historical-data (accessed 21 August 2018).
	\item[35]  Quandl, Quandl.Com. (2018). https://www.quandl.com/data/CHRIS/CME\_GC1-Gold-Futures-Continuous-Contract-1-GC1-Front-Month (accessed 21 August 2018). 
	
	\item[36]  Quandl, Quandl.Com. (2018). https://www.quandl.com/data/ODA/POILWTI\_USD-WTI-Crude-Oil-Price (accessed 21 August 2018). 
	
	\item[37] USD INR | US Dollar Indian Rupee - Investing.com India, Investing.Com India. (2018). https://in.investing.com/currencies/usd-inr (accessed 21 August 2018).
	\item[38] C. Granger, Investigating Causal Relations by Econometric Models and Cross-spectral Methods, Econometrica. 37 (1969) 424. doi:10.2307/1912791.
	\item[39] D. Dickey, W. Fuller, Distribution of the Estimators for Autoregressive Time Series with a Unit Root, Journal Of The American Statistical Association. 74 (1979) 427-431. doi:10.1080/01621459.1979.10482531.
	\item[40] H. Bierens, A Companion to Theoretical Econometrics, Blackwell Publishing Ltd, 2003, Pages:610-633.
	\item[41] A. Haar, Zur Theorie der orthogonalen Funktionensysteme, Mathematische Annalen. 69 (1910) 331-371. doi:10.1007/bf01456326.
	\item[42] L. Sjaastad, The price of gold and the exchange rates: Once again, Resources Policy. 33 (2008) 118-124. doi:10.1016/j.resourpol.2007.10.002.
	\item[43] A. Koutsoyiannis, A short-run pricing model for a speculative asset, tested with data from the gold bullion market, Applied Economics. 15 (1983) 563-581. doi:10.1080/00036848300000037.
	\item[44] R. Pindyck, Investments of uncertain cost, Journal Of Financial Economics. 34 (1993) 53-76. doi:10.1016/0304-405x(93)90040-i.
	\item[45] R. Laurent, Is there a role for gold in monetary policy? Economic Perspectives., The Federal Reserve Bank Of Chicago. 13 (1994) 2-14.
	\item[46] S. Mahdavi, S. Zhou, Gold and commodity prices as leading indicators of inflation: Tests of long-run relationship and predictive performance, Journal Of Economics And Business. 49 (1997) 475-489. doi:10.1016/s0148-6195(97)00034-9.
	\item[47] G. Moore, Analysis, Challenge. 33 (1990) 52-56. doi:10.1080/05775132.1990.11471444.
	\item[48] S. Johansen, Estimation and Hypothesis Testing of Cointegration Vectors in Gaussian Vector Autoregressive Models, Econometrica. 59 (1991) 1551. doi:10.2307/2938278.
	\item[49] A. BHUNIA, COINTEGRATION AND CAUSAL RELATIONSHIP AMONG CRUDE PRICE, DOMESTIC GOLD PRICE AND FINANCIAL VARIABLES- AN EVIDENCE OF BSE AND NSE, Journal Of Contemporary Issues In Business Research. 2 (2013) 1-10.
	\item[50] N. Ricker, WAVELET CONTRACTION, WAVELET EXPANSION, AND THE CONTROL OF SEISMIC RESOLUTION, GEOPHYSICS. 18 (1953) 769-792. doi:10.1190/1.1437927.
	\item[51] T. Qian, M. Vai, Y. Xu, Wavelet Analysis and Applications, Birkhaauser Verlag, Basel, 2007.
	\item[52] S. Mallat, A wavelet tour of signal processing, Academic Press, San Diego, 1998.
	\item[53] A. Akansu, R. Haddad, Multiresolution signal decomposition, Academic Press, San Diego, CA, 1992.
	\item[54] E. Cohen, A. Walden, A Statistical Study of Temporally Smoothed Wavelet Coherence, IEEE Transactions On Signal Processing. 58 (2010) 2964-2973. doi:10.1109/tsp.2010.2043139.
	\item[55] L. Aguiar-Conraria, M. Soares, THE CONTINUOUS WAVELET TRANSFORM: MOVING BEYOND UNI- AND BIVARIATE ANALYSIS, Journal Of Economic Surveys. 28 (2013) 344-375. doi:10.1111/joes.12012.
	\item[56] P. PHILLIPS, P. PERRON, Testing for a unit root in time series regression, Biometrika. 75 (1988) 335-346. doi:10.1093/biomet/75.2.335.
	\item[57] A. Bhargava, On the Theory of Testing for Unit Roots in Observed Time Series, The Review Of Economic Studies. 53 (1986) 369. doi:10.2307/2297634.
	\item[58] A. Jain, S. Ghosh, Dynamics of global oil prices, exchange rate and precious metal prices in India, Resources Policy. 38 (2013) 88-93. doi:10.1016/j.resourpol.2012.10.001.
	\item[59] A. Tiwari, I. Sahadudheen, Understanding the nexus between oil and gold, Resources Policy. 46 (2015) 85-91. doi:10.1016/j.resourpol.2015.09.003.
	\item[60] E. Bouri, A. Jain, P. Biswal, D. Roubaud, Cointegration and nonlinear causality amongst gold, oil, and the Indian stock market: Evidence from implied volatility indices, Resources Policy. 52 (2017) 201-206. doi:10.1016/j.resourpol.2017.03.003.
	\item[61] A. Jain, P. Biswal, Dynamic linkages among oil price, gold price, exchange rate, and stock market in India, Resources Policy. 49 (2016) 179-185. doi:10.1016/j.resourpol.2016.06.001.
	\item[62] W. Hu, B. Si, Technical note: Multiple wavelet coherence for untangling scale-specific andlocalized multivariate relationships in geosciences, Hydrology And Earth System Sciences. 20 (2016) 3183-3191. doi:10.5194/hess-20-3183-2016.
\end{itemize}

\end{document}